\documentclass[journal]{IEEEtran}
\usepackage[cmex10]{amsmath}
\usepackage{array}
\usepackage{mdwmath}
\usepackage{mdwtab}
\usepackage{eqparbox}

\usepackage{bm}
\usepackage{multicol}
\usepackage{multirow}
\usepackage{hhline} 
\usepackage{rotating}
\usepackage{mathtools,amssymb,lipsum, nccmath}

\usepackage{color, soul}
\usepackage{framed}
\usepackage{times,amsmath,epsfig}
\usepackage{amsmath}
\usepackage{amscd}
\usepackage{amssymb}
\usepackage{graphicx}
\usepackage[tight,footnotesize]{subfigure}
\usepackage{epstopdf}
\usepackage{pdfpages}

\usepackage{hyperref}
\usepackage{tablefootnote}
\usepackage{tabularx, booktabs}
\usepackage{cancel}
\usepackage[normalem]{ulem}

\usepackage{cuted}

\usepackage[numbers,sort&compress]{natbib}
\usepackage[linesnumbered,ruled]{algorithm2e}
\SetKwRepeat{Do}{do}{while}

\def\be{ \begin{eqnarray} }
\def\ee{ \end{eqnarray} }
\def\bit { \begin{item} }
\def\eit { \end{item} }
\def\bnum { \begin{enumerate} }
\def\enum { \end{enumerate} }

\begin{document}
%
%
%

\title{\huge Multi-population Differential Evolution for RSS based  Cooperative Localization in Wireless Sensor Networks with Limited Communication Range}

\author{Lismer Andres Caceres Najarro,
	Iickho Song,~\IEEEmembership{Fellow,~IEEE},
        Muhammad Salman, and 
	Kiseon Kim,~\IEEEmembership{Senior Member,~IEEE}
     \thanks{L. A. Caceres Najarro is with the Department of Computer Science of Chosun University, 
       Gwangju 61452
     Republic of Korea. Email: andrescn@chosun.ac.kr}
        \thanks{K. Kim is
		with the School of Electrical Engineering and Computer Science, Gwangju
		Institute of Science and Technology,
		Gwangju 61005 Republic of Korea. Email: kskim@gist.ac.kr}
        \thanks{M.Salman is with College of Electrical \& Mechanical Engineering, National University of Sciences and Technology, Islamabad, Pakistan. Email: msalman@ceme.nust.edu.pk}
	\thanks{I. Song is with the School of Electrical Engineering, Korea Advanced
		Institute of Science and Technology, Daejeon 34141 Republic of Korea. Email: i.song@ieee.org}
	\thanks{
		This work was supported by
		the National Research Foundation of Korea
		under Grant NRF-2021R1I1A1A01041257,
		%
		in part by the project titled “Development of Automatic Identification Monitoring System for Fishing Gears”, funded by the Ministry of Oceans and Fisheries, Korea,
		and in part by
		the University of Science and Technology of China
		Visiting Professor International under
		Grant No. 2022BVT04, and by 
            the research fund from Chosun University, 2024,
		for which the authors wish to express their appreciation.
	}
}


%



\maketitle

\setlength{\baselineskip}{1\baselineskip}

\begin{abstract}

%
This paper presents a novel approach to deal with the cooperative localization problem in wireless sensor networks based on received signal strength measurements.
%
%
%
In cooperative scenarios,
the cost function of the localization problem becomes increasingly nonlinear and nonconvex due to the heightened interaction between sensor nodes, making the estimation of the positions of the target nodes more challenging.
%
%
Although
most of existing
cooperative localization algorithms assure acceptable localization accuracy,
their computational complexity increases dramatically, which may restrict their applicability.
%
To
reduce the computational complexity and provide competitive localization accuracy at the same time,
we propose a localization algorithm based on the differential evolution with multiple populations, opposite-based learning, redirection, and
anchoring.
In this work, the cooperative localization cost function is split into several simpler cost functions,
each of which accounts
only for one individual target node. Then, each cost function is solved by a dedicated population of the proposed algorithm.
%
%
In addition,
an enhanced version of the proposed algorithm which incorporates the population midpoint scheme for further improvement in the localization accuracy is devised.
Simulation results demonstrate that the proposed algorithms provide comparative localization accuracy with much lower computational complexity compared with the state-of-the-art algorithms.

\end{abstract}

\begin{IEEEkeywords}
Centralized, cooperative, localization, multiple population, differential evolution,
received signal strength (RSS).
\end{IEEEkeywords}

%
\IEEEpeerreviewmaketitle

\section{Introduction}\label{sec_Intro}
\IEEEPARstart
%
%
{T }{}he attention to wireless sensor networks (WSNs) has been increasing in recent years.
This is mainly because WSNs allow the installation of a large number of small, low-power sensor nodes
in
an
area of interest (AOI)
for a wide range of applications such as
monitoring, tracking,
smart cities, autonomous vehicles,
and smart farming \cite{caceres2022fundamental}.
%
%
Furthermore,
it is expected that the applicability of WSNs will continuously increase due to the rise of new technologies such as 6G \cite{chowdhury2020_6G}
and
cheaper
prices of sensor nodes \cite{Online_Yinbiao_WSN_2019}.
Normally,
WSNs consist of several
anchor nodes (ANs) and target nodes (TNs):
the ANs can be static or dynamic and are usually aware of their own positions. In contrast, the TNs are sensor nodes with an unknown position that need to be estimated.
%
%
%
%

The estimation of
the locations of the
TNs is of great importance for any application of WSNs.
In the literature there exist several techniques for estimating the unknown positions of the TNs,
which can be grouped into
centralized or decentralized approaches \cite{caceres2022fundamental}.
%
Although we focus our attention on centralized approaches in this work, the proposed algorithm can be easily implemented in a distributed manner also.
The locations of TNs are commonly found via (a combination of) signal related measurements such as received signal strength (RSS) \cite{jour_Zanella_RSS_Measur_2016}, time of arrival (TOA) \cite{wu2023multistatic}, angle of arrival (AOA) \cite{shih2023angle}, and
channel state information (CSI) \cite{Tian_CSI_indoor_2019}.
Among these common measurements, RSS is preferred due to the fact that most of the commercial
sensor nodes readily provide the RSS information, and thus additional hardware/software is not required
as in the case of the other measurements.
This allows an easier implementation of the RSS-based localization with low hardware complexity and low cost.
For these reasons, in this work, we utilize RSS measurements to find the locations of several TNs in a cooperative manner.

In contrast to the
noncooperative localization, the cooperative localization uses signal related measurements obtained in both ANs and TNs.
This increases the information about the TNs, which leads to an improvement in the localization accuracy, especially in sparse WSNs \cite{caceres2022fundamental}.
The improvement in the localization accuracy
is possible
at the cost of
increased
complexity of the maximum likelihood (ML)
cost function
\cite{book_Estimation_opt}.
Such highly complex ML cost function results in
a more challenging optimization problem that cannot be handled by
standard optimization techniques.
A possible way to solve this problem is to employ iterative techniques.
Although iterative techniques can be used for solving the cooperative localization problem, they
generally require initial guess points that usually restrict their applicability.
%
%
%
To avoid this requirement and guarantee convergence
in some extent,
convex optimization \cite{book_Boyd} has recently attracted considerable attention for solving the cooperative localization problem.

Techniques based on semidefinite programming (SDP) and second order cone programming (SOCP) have been shown to outperform several existing techniques based on the first/second derivative and linearization methods.
%
%
%
%
For instance, under the assumption of low noise, the cooperative localization problem was relaxed using convex relaxation techniques based on SDP \cite{jour_Tomic}.
Similar relaxation was adopted in \cite{jour_Chang_2018} based on SOCP when the cooperative localization problem was formulated under weighted least square criterion.
%
In addition to the approaches in
\cite{jour_Tomic} and \cite{jour_Chang_2018},
the log-normal RSS measurement model
was transformed into an equivalent multiplicative model in \cite{jour_Z_Wang_2019}. The cooperative ML estimator was approximated by a nonconvex estimator using the relative error criterion \cite{jour_Chen_2010}, which was then relaxed based on SDP techniques.
Recently,
biased RSS measurements were considered for formulating the cooperative localization problem, which was then relaxed based on SDP \cite{wang2022SDP_biased_RSS}.
More recently,
the cooperative localization cost function was converted into invariant convex function for reducing the computational complexity, which was then solved by the gradient descent algorithm \cite{mukhopadhyay2022invex}.
Although
the approach in \cite{mukhopadhyay2022invex} is computationally more efficient, it provides a lower localization accuracy than that in \cite{wang2022SDP_biased_RSS}.
%
%
%
Certainly, the algorithms based on SDP and SOCP provide a reasonable localization accuracy, but,
unfortunately, they suffer from a high computational complexity restricting their applicability.

In this paper,
to reduce the computational complexity and provide a competitive localization accuracy at the same time,
we investigate an alternative optimization technique based on a novel variant of the differential evolution (DE) \cite{jour_Storn}.
%
Unlike
the aforementioned cooperative localization algorithms,
we
first
reformulate the cooperative localization problem under the multi-objective optimization framework.
To do so, the cost function of the
ML
formulation of the cooperative localization problem is cast into several objective functions that are optimized separately.
%
%
Although evolutionary algorithms with a single population may solve the cooperative localization problem,
a
multi-population evolutionary algorithm based on the DE is proposed to avoid the curse of dimensionality \cite{chen2022CurseDimen}.
The proposed algorithm also takes advantage of the opposition-based learning (OBL) \cite{conf_Tizhoosh_Oppo_teo_2005, jour_Seif_Oppo_appli_2015} and adaptive redirection \cite{caceres2020DEOR} for better search capability and redirection of the individuals, respectively.
Additionally,
we exploit the connectivity information inferred from
the interaction between sensor nodes for a better initialization and guiding of the individuals towards the positions of the TNs.
%
To be precise,
the connectivity information is exploited for generating the initial populations by generating pseudo-random individuals that are close to the ANs.
Furthermore,
a new process called anchoring is devised for guiding the individuals through the evolution process of all the populations.

%

The rest of this paper is organized as follows. In Section \ref{sec_Model}, we present the RSS measurement model, the ML estimator, and the multi-objective framework for the cooperative localization problem. In Section \ref{sec_Approach}, the proposed approach is described in detail.
The simulation results for several localization scenarios are presented in Section \ref{sec_Simulation}, where the proposed approach is compared with the state-of-the-art algorithms. The computational complexity is also analyzed in this section. Finally, concluding remarks are provided in Section \ref{sec_Conclusion}.


\section{System Model} \label{sec_Model}

This section introduces the cooperative localization problem based on RSS measurements,
%
%
and
presents the reformulation of the cooperative localization problem
under the umbrella of multi-objective optimization framework.

\subsection{Cooperative Localization}

Let us consider a centralized WSN in the two-dimensional space $\mathbb{R}^2$
with $N$ ANs and $M$ TNs having the same communication, where $M \gg N$.
The extension to the three dimensional space should be straightforward.
We say that a pair of two sensors with positions
$\bm{p}_1$ and $\bm{p}_2$  in $\mathbb{R}^2$
are directly connected or have a direct connection if only if
\be
\label{eq_direct_connection}
\left\| \bm{p}_1 - \bm{p}_2\right\| \le R,
\ee
where $\|\cdot\|$ and $R$ denote the $\mathcal{L}_2$ norm and the communication range of the sensor nodes, respectively.
If (\ref{eq_direct_connection}) does not hold, then we say the sensor nodes are indirectly connected or have an indirect connection.
In an indirect connection, the two sensor nodes require
relay
node(s) to communicate each other since they are not within their communication range.
We assume,
due to a hardware limitation or limited communication range,
it is not possible to have a fully connected network:
in other words,
only part of the TNs are directly connected to the ANs.
%
%
In this situation,
the cooperation between sensor nodes
is
necessary
and
is required to
compensate for the deficiency in the number of ANs for
an
accurate estimation of the positions of TNs.
This means that
the TNs will also conduct RSS measurements and participate actively as pseudo ANs in the localization process.
%
%

The positions of the $n$-th AN and the $m$-th TN are denoted by
$\bm{s}_n = {\left[ s_{n,1} , s_{n,2} \right]}, n \in \mathcal{N}$
and
$\bm{x}_m = {\left[ x_{m,1} , x_{m,2} \right]}, m\in \mathcal{M}$,
where $\mathcal{N}=\{1,2,\cdots, N\}$ and
$\mathcal{M}=\{N+1,N+2,\cdots, N+M\}$ represent the set of indices of the ANs and TNs,
respectively.
%
%
To find the positions of TNs in a cooperative manner, RSS measurements 
are
conducted in both the TNs and ANs.
Note that the RSS measurements can only be taken between sensor nodes with positions satisfying (\ref{eq_direct_connection}).
%
Thus, we define two sets
%
%
%
$\mathcal{A}_m=\{n: \left\| \bm{s}_n - \bm{x}_m \right\| \le R,\quad n \in \mathcal{N}\}$
and
%
$\mathcal{B}_m=\{k: \left\| \bm{x}_k - \bm{x}_m\right\|  \le R,\quad k \in \mathcal{M} \}$
that contain the indices of the anchor nodes and target nodes, respectively, directly connected to the $m$-th TN
for $ m \in \mathcal{M}$.
%
%
The
information contained in the sets $\mathcal{A}_m$ and $\mathcal{B}_m$ will be used later in Section \ref{sec_Approach}
in
the generation of the initial population and anchoring processes of the proposed technique.
%

%

The path loss between the $n$-th AN (TN) and the $m$-th TN in decibel
\cite{jour_Tomic,mukhopadhyay2022invex} can be formulated as
\be\label{eq_PL}
L_{nm} = L_0 + 10\gamma {\log _{10}}d_{nm} + v_{nm}
%
\ee
%
for $m\in \mathcal{M}$ and $n \in \mathcal{A}_m \cup \mathcal{B}_m$,
where
$L_0$ denotes the path loss at the reference distance 
assumed to be $1$ m as in the literature \cite{jour_Chang_2018, jour_Z_Wang_2019, mukhopadhyay2022invex},
$\gamma$ is the path loss exponent,
$v_{nm}$ represents the log-shadowing noise 
modeled as a zero-mean Gaussian random variable with variance $\sigma _{{v_{nm}}}^2$,
and
$d_{nm}$ is the Euclidean distance between the $n$-th and $m$-th sensor nodes, i.e.,
for $m \in \mathcal{M}$, $d_{nm}= \left\| \bm{s}_n - \bm{x}_m \right\|$ when
$n \in \mathcal{N}$
and
$d_{nm}= \left\| \bm{x}_n - \bm{x}_m \right\|$ when
$ n \in \mathcal{M}$.

Let us denote by
$\bm{X}=\left[\bm{x}_{N+1}^T, \bm{x}_{N+2}^T, \cdots, \bm{x}_{N+M}^T  \right]$
$\in  \mathbb{R}^{2 \times M}$
the
vector
of
all the unknown positions of the TNs. Then, from (\ref{eq_PL}) the ML estimator of $\bm{X}$ for cooperative RSS based localization can be obtained as
%
%
%
\begin{IEEEeqnarray}{rCl}\label{eq_cost_func_Cooperative}
{\hat{\bm{X}}} =
\mathop {\arg \min }\limits_{\bm{X}} \sum\limits_{m \in \mathcal{M}} {\sum\limits_{n \in \mathcal{A}_m\cup \mathcal{B}_m} {\frac{{\left({P_{nm} - 10\gamma {{\log }_{10}}d_{nm}} \right) } ^2}{\sigma _{{v_{nm}}}^2}}},
\end{IEEEeqnarray}
where $P_{nm}=L_{nm} -L_0$.
The cost function of (\ref{eq_cost_func_Cooperative}) is highly nonlinear, multimodal, and nonconvex:
this
is
the reason why most of the state-of-the-art algorithms solve an approximation of the ML
formulation.

\subsection{Multiobjective Formulation}
%
%
%
The cooperative localization problem can be viewed as a dynamic optimization problem,
and
can be reformulated under the  multi-objective optimization framework \cite{jour_Feng_MOP_2017}.
%
To do so, we first note that (\ref{eq_cost_func_Cooperative}) can be rewritten as
%
%
%
\begin{IEEEeqnarray}{rCl}\label{eq_ML}
	\hat {\bm{X}} =\mathop {\arg \min }\limits_{\bm{X}} {\sum\limits_{m = N+1}^{N+M} {{f_m}\left( {\bm x }_m\right)}},
\end{IEEEeqnarray}
where
\begin{IEEEeqnarray}{rCl}\label{eq_cost_func}
		f_m({\bm{x}_m}) =
{\sum\limits_{n \in \mathcal{A}_m\cup \mathcal{B}_m} {\frac{1}{\sigma _{{v_{nm}}}^2}}{\left({P_{nm} - 10\gamma {{\log }_{10}}d_{nm}} \right) } ^2}
\end{IEEEeqnarray}
is the cost function corresponding to the
$m$-th TN for
$m\in \mathcal{M}$.
%
%
To align with the multi-objective optimization framework, each TN's cost function is treated as a separate objective, allowing simultaneous minimization of individual cost function as
\small
\be\label{eq_MOP}
\begin{split}
	\bm{\hat{X}}=\mathop {\arg \min }\limits_{\bm{X}}\{f_{N+1}\left(\bm{x}_{N+1}\right), f_{N+2}\left(\bm{x}_{N+2}\right), \\
 \cdots, f_{N+M}\left(\bm{x}_{N+M}\right)\}.
\end{split}
\ee
\normalsize
This strategy is expected to enhance the localization accuracy by optimizing each position of the TNs while considering the interactions among them.


%
%
%
%

\section{Proposed Localization Algorithm}\label{sec_Approach}

For solving (\ref{eq_MOP}), we propose a multiple population DE combined with OBL, adaptive redirection, and an additional process called anchoring.
The OBL
in the proposed method,
is used in the initialization process in order to generate better individuals closer to the global optimum.
In the meantime,
the adaptive redirection is to assure that the individuals
are located
inside the solution space
\be
\bm{Z} = \left\{ \left( z_1 , z_2 \right): a_1 \le z_1 \le b_1,
a_2 \le z_2 \le b_2 \right\},
\ee
also called the
area of interest (AOI),
where $a_d$ and $b_d$ denote, respectively, the lower and upper bound of the AOI for $d=1,2$.
The process anchoring
guides
the individuals through the evolution process toward the global optimum
by using
the connectivity information contained in the sets
$\mathcal{A}_m$ and $\mathcal{B}_m$.
%
%
%


\begin{figure}[!t]
	\centering
	\includegraphics[width=\linewidth]{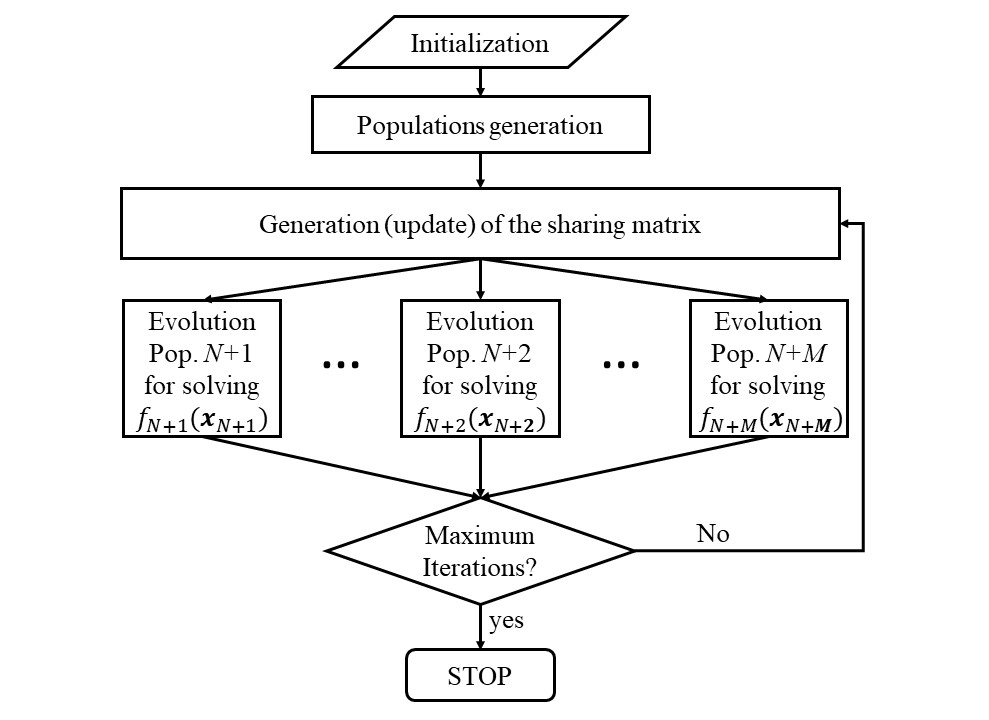} 
	\caption{Flow chart of the proposed algorithm for the cooperative localization in WSNs.}
	\label{fig_MDEOR_flowchart}
\end{figure}
%


Fig. \ref{fig_MDEOR_flowchart} shows the flow chart of the main components of the proposed algorithm.
%
The proposed algorithm generates as many
populations
as
the number $M$ of the TNs.
The $m$-th population is dedicated to solve the cost function
$f_m\left(\bm{x}_m\right)$ for
$m\in \mathcal{M}$.
%
%
%
The sharing matrix serves as a crucial link between the populations,
storing
and updating
the estimated positions of all TNs at each generation.
This matrix facilitates the sharing of the estimates of the positions of the TNs among the populations, ensuring collaborative information exchange.
More details on
the sharing matrix will be
described
later when we
present
the pseudo code of the proposed algorithm.

%


\subsection{Processes of the Proposed Algorithm}\label{subsec_proposed algorithm_steps}
Commonly, standard evolutionary algorithms such as the DE start the generation of individuals randomly.
In the proposed algorithm, on the other hand, a pseudo random initialization by using connectivity information between ANs and TNs
is devised.
%
%
%
%
%
In essence, if a 
TN is directly connected to one or more ANs, the individuals of the 
corresponding population will be
generated by taking into account the positions of the ANs
to which
the 
TN is directly connected.
%
By doing so,
it is expected that
the initial population will be closer to the global minima, resulting in a faster convergence of the proposed algorithm.


%

Let us denote the $m$-th population,
a collection of $L$ individuals,
in the $g$-th generation by
$\left\{ {\bm{I}_l^{g,m}} \right\}_{l = 1}^L$
for $g=1,2, \cdots, G$ and $m=1,2, \cdots, M$.
Each individual $\bm{I}_l^{g,m}$ is a two-dimensional vector $\left[I_{l,1}^{g,m},I_{l,2}^{g,m} \right]$, with $I_{l,1}^{g,m}$  and $I_{l,2}^{g,m}$ called genes,
and is a candidate estimate of the $m$-th unknown position ${\bm{x}_m}$.

\vspace{2mm}
1) \textit{Generation of initial populations:}
The initial $m$-th population $\left\{ {\bm{I}_l^{1,m}} \right\}_{l = 1}^L$ is selected taking the best $L$
vectors
among $\left\{ {\bm{I}_l^{+,m}} \right\}_{l = 1}^L$ and $\left\{ {\bm{I}_l^{-,m}} \right\}_{l = 1}^L$
for $m=1,2,\cdots,M$,
which are
created
as pseudo-random vectors 
%
%
%
%
\be
\label {eq_Ini_pop}
{I}_{l,d}^{+,m} = \left\{ {\begin{array}{*{20}{c}}
{a_d+r\left(b_d -a_d\right),{\quad \rm{if}\quad }{\delta} = 0},\\
{ R_d + \frac{1}{|\mathcal{A}_m|}{{\sum\limits_{n \in \mathcal{A}_m} {{s_{n,d}}} }},{\quad  \rm{if} \quad \rm{\delta=1}}},
\end{array}} \right.
\ee
and their opposite
%
%
%
%
\be
\label {eq_Ini_pop_oppo}
{I}_{l,d}^{-,m} = \left\{ {\begin{array}{*{20}{c}}
{a_d+b_d - {I}_{l,d}^{+,m},{\quad \rm{if}\quad }{\delta} = 0},\\
{  \frac{2}{|\mathcal{A}_m|}{{\sum\limits_{n \in \mathcal{A}_m} {{s_{n,d}}} }}-{I}_{l,d}^{+,m},{\quad \rm{if}\quad  \rm{\delta=1}}},
\end{array}} \right.
\ee
using the OBL scheme \cite{conf_Tizhoosh_Oppo_teo_2005}, respectively,
where
$\delta=1$
when the $m$-th TN is directly connected to one or more ANs and $\delta=0$ otherwise,
$R_d=\frac{rR}{ |\mathcal{A}_m|} \cos(\theta)$ and
$R_d=\frac{rR}{ |\mathcal{A}_m|} \sin(\theta)$
%
%
when $d=1$ and $d=2$, respectively,
$r$
is the product of two
uniformly distributed random variables on $\left[0, 1\right]$,
$\theta $ is a random variable uniformly distributed on $\left[0, 2\pi\right]$,
and
%
$| \cdot |$ denotes the cardinality of a set. 

Once the initial populations are generated,
five processes, namely,
mutation, crossover, anchoring, adaptive redirection, and selection are conducted recursively
for each of the $M$ populations.


\vspace{2mm}

2) \textit{Mutation:}
The $l$-th mutant of the $m$-th population is generated from the combination of three individuals at the $g$-th generation as
\begin{IEEEeqnarray}{rCl} \label {eq_Mutation}
\tilde{\bm{I}}_l^{g,m} = \bm{I}_k^{g,m} + \alpha \left( \bm{I}_p^{g,m} - \bm{I}_q^{g,m} \right).
\end{IEEEeqnarray}
In (\ref{eq_Mutation}) $\alpha$ is the scaling factor determining the diversity and convergence speed with the value normally between 0.4 and 1 \cite{jour_Ronkkone}.
The indices $k$, $p$, and $q$ of the three individuals are randomly selected in such a way that they are different from each other.

\vspace{2mm}
3) \textit{Crossover:} Once the mutation process is completed, a trial
individual
$\breve{\bm{I}}_l^{g,m}=\left[ \breve{I}_{l,1}^{g,m} , \breve{I}_{l,2}^{g,m} \right]$
is created as
\be
\label {eq_Crossover}
\breve{I}_{l,d}^{g,m} \ = \ \left\{ \begin{array}{ll}
\tilde I_{l,d}^{g,m}, & \textrm{ if } c_l \le p_C, \\ 
I_{l,d}^{g,m}, & \textrm{ if } c_l > p_C
\end{array} \right.
\ee
by mating the mutated vector $\tilde{\bm{I}}_l^{g,m}$ and the target vector ${\bm{I}}_l^{g,m}$ for $l=1,2, \cdots, L$,
where $c_l$ is a uniformly distributed random variable
on
$\left[0, 1\right]$ and $p_C \in \left[0, 1\right]$ represents the crossover probability.

\vspace{2mm}
4) \textit{Anchoring:}
In an evolutionary algorithm, the trial individuals are commonly blindly generated:
i.e.,
a trial individual is generated without any information.
In contrast, in the proposed algorithm,
the connectivity information contained in the sets $\{A_m\}$ and $\{B_m\}$ is used to guide the individuals toward the positions of TNs.
The guidance is provided by anchoring the individuals with respect to the positions of the ANs and/or TNs to
which the $m$-th TN is directly connected.
Specifically, the $m$-th anchored population $ \{\dddot{\bm{I}}_l^{g,m} =\left[\dddot{{I}}_{l,1}^{g,m},\dddot{{I}}_{l,2}^{g,m} \right]\}_{l=1}^L$ is obtained from the trial population $\{\breve{\bm{I}}_l^{g,m}\}_{l=1}^L$ as
\be
\label {eq_Anchoring_general}
%
%
\dddot{{I}}_{l,d}^{g,m} = \left\{ {\begin{array}{*{20}{c}}
		{\breve{{I}}_{l,d}^{g,m},{\quad \rm{if}\quad }{C_N} = {C_M} = 0},\\
		{\frac{{\dot{{I}}_{l,d}^{g,m} \cdot {C_N} + \ddot{{I}}_{l,d}^{g,m} \cdot {C_M}}}{{ {C_N} + {C_M}}},{\quad  \rm{otherwise}}},
\end{array}} \right.
\ee
where $C_N$ and $C_M$ are binary parameters:
$C_N=0$ if the trial individual $\breve{\bm{I}}_l^{g,m}$ satisfy
(\ref{eq_direct_connection}) for direct connection with respect to the positions of the ANs the $m$-th TN is connected to, otherwise $C_N=1$.
Similarly, $C_M=0$ if the trial individual $\breve{\bm{I}}_l^{g,m}$ satisfy the condition (\ref{eq_direct_connection}) for direct connection with respect to the positions of the TNs the $m$-th TN is connected to, otherwise $C_M=1$.
The terms
%
%
\begin{IEEEeqnarray}{rCl} \label {eq_Anchoring_ANTN}
\dot I_{l,d}^{g,m} = R_d + \frac{1}{|\mathcal{A}_m|}{\sum\limits_{n \in \mathcal{A}_m} {{s_{n,d}}} }
\end{IEEEeqnarray}
%
%
and
\begin{IEEEeqnarray}{rCl} \label {eq_Anchoring_TNTN}
\ddot I_{l,d}^{g,m} = R_d + \frac{1}{|\mathcal{B}_m|}{\sum\limits_{k \in \mathcal{B}_m} {\hat x_{k,d}^{g - 1}} }
\end{IEEEeqnarray}
represent anchored vectors relative to the positions of the ANs and TNs,
respectively.
%
%
In addition,
$\hat{\bm{x}}_k^{g - 1}={\left[ \hat{x}_{k,1}^{g-1} , \hat{x}_{k,2}^{g-1}\right]}$ denotes the best estimate of the $k$-th TN  with direct connection to the $m$-th TN, obtained at the ($g-1$)-th generation.
%

\vspace{2mm}
5) \textit{Adaptive redirection:}
Due to the random generation of the populations and the mutation process,
some individuals may be out of the solution space $\bm Z$ affecting negatively the evolution process.
To avoid this situation,
the adaptive redirection process introduced in \cite{caceres2020DEOR} is employed.
The adaptive redirection process uses efficiently the information contained in the population to guide the individuals.
To be specific,
to assure that the genes of individuals are always within the solution space
$\bm{Z}$, 
the redirected population $\left \{ \cancel{\bm{I}}_{l}^{g,m} =
\left[ \cancel{I}_{l,1}^{g,m} , \cancel{I}_{l,2}^{g,m} \right] \right \}_{l=1}^{L}$
is obtained as
\be \label{eq_redi}
\cancel I_{l,d}^{g,m} = \left\{ \begin{array}{ll}
\dddot{I} _{l,d}^{g,m} , \ \textrm{ if } a_d \le \dddot{I} _{l,d}^g \le b_d ,\\
\beta_d \left( {b_d^{g,m} - a_d^{g,m}} \right) + a_d^{g,m}\ , \\
\qquad \textrm{ if } \dddot{I} _{l,d}^{g,m} < a_d\textrm{ or }  \dddot{I} _{l,d}^{g,m}  > b_d
\end{array} \right.
\ee
for $d=1,2$
from the anchored population $\left \{ \dddot{\bm{I}}_{l}^{g,m} \right \}_{l=1}^{L}$, where
$\beta_d$
is a uniformly distributed random variable
on
$\left[0, 1\right]$
and
%
%
\be
a_d^{g,m} = \mathop {\min }\limits_{I_{l,d}^{g,m}}
\left( \left \{ I_{l,d}^{g,m} \right \}_{l=1}^L \right)
\ee
and
%
%
\be
b_d^{g,m} = \mathop { \max }\limits_{I_{l,d}^{g,m}}
\left( \left \{I_{l,d}^{g,m} \right \}_{l=1}^L  \right)
\ee
denote the lower and upper bounds, respectively, of the area of
redirection at the $g$-th generation of the $m$-th population.


\vspace{2mm}
6) \textit{Selection:}
At each iteration $g$, the proposed algorithm selects the individual with the lowest fitness value as
\begin{IEEEeqnarray}{rCl} \label {eq_Selection}
\bm{I}_{l,m}^{g + 1} = \mathop {\arg \min }
\limits_{\bm{I}_l^{g,m}, \cancel {\bm{I}}_l^{g,m}}
\left \{ f_m\left( \bm{I}_l^{g,m} \right) , f_m\left( \cancel {\bm{I}}_l^{g,m} \right) \right\},
\end{IEEEeqnarray}
among the target individual $\bm{I}_l^{g,m} $ and the redirected individual ${\cancel {\bm{I}}}_l^{g,m}$.
Note that the populations are updated only if the redirected individual is better than the target individual.

\vspace{1ex}

After all $M$ populations evolve during $G$ generations by passing through the five processes explained above, the best individual of the $m$-th population is selected as the best estimation of the $m$-th TN as
\begin{IEEEeqnarray}{rCl} \label {eq_Fin_proposed algorithm}
\hat{\bm{x}}_m = \mathop {\arg \min }\limits_{\bm{I}_l^{G,m}} \left \{
f_m \left ( \bm{I}_l^{G,m} \right ) \right\}_{l=1}^{L}
\end{IEEEeqnarray}
for
$m\in \mathcal{M}$.
Then, the estimate of the $M$ TNs $\hat{\bm{X}}= \left[ \hat{\bm{x}}_{N+1}^T, \hat{\bm{x}}_{N+2}^T, \cdots,\hat{\bm{x}}_{N+M}^T \right]$ is presented.

\subsection{Enhancement of the Proposed Algorithm}
Here,
we incorporate the population midpoint scheme \cite{arabas2017improving}, which was demonstrated to improve the performance of the standard DE in a simple and elegant way \cite{caceres2022midpoint}.
The population midpoint scheme
creates a midpoint individual 
$\bm{\bar{I}}^g \!=\! [\bar{I}^g_1,\bar{I}^g_2]$ by
calculating the average over a
population
$\left\{ {\bm{I}_l^{g}} \right\}_{l = 1}^L$
as

%
\begin{equation}
	\bm{\bar{I}}^g = \cfrac{1}{L} \sum_{l=1}^L \bm{I}_l^g,
	\label{eq_mid_point}
\end{equation}
where $\bm{I}_l^g=\left[I_{l,1}^{g},I_{l,2}^{g} \right]$ represents an individual in the population.

%

The midpoint individual has been proven statistically that it is better than any individual in the population \cite{arabas2017improving}.
Although the midpoint scheme can be used at any stage of the evolution process, it has been demonstrated in \cite{caceres2022midpoint} that a consistent improvement can be obtained when the population midpoint scheme is employed after $G$ generations.
Consequently,
we employ the midpoint scheme
at the end of the evolution process
for
each of the populations.
%
In other words,
the estimated position of the $m$-th TN
is
obtained as

\begin{equation}
	\hat{\bm{x}}_{m} = \cfrac{1}{L} \sum_{l=1}^{L} \bm{I}^{G,m}_l.
	\label{eq_midpoint_TNs}
\end{equation}
Note that in contrast to (\ref{eq_Fin_proposed algorithm}), instead of taking the best individual, we take the individual obtained through the population midpoint scheme as the estimate of the $m$-th TN.


\vspace{1ex}

The pseudo code of the proposed algorithm is presented in Algorithm \ref{algo1}.
Here,
%
$\hat{\bm{X}}^{g}= \left[ \hat{\bm{x}}_{N+1}^g, \hat{\bm{x}}_{N+2}^g, \cdots, \hat{\bm{x}}_{N+M}^g \right]$
is 
the sharing matrix in which the estimates of the positions of all TNs at the $g$-th generation are stored and updated.
This matrix is used for evaluating the cost function and anchoring (\ref{eq_Anchoring_general}) the individuals in each population.
Note that this matrix is fully updated generation by generation, and
acts as
a
link between the populations.

\begin{algorithm}[t!]\label{algo1}
    \SetKwInOut{Input}{Input}
    \SetKwInOut{Output}{Output}

    \Input{ML cost function (\ref{eq_cost_func}), RSS measurements, number $N$ of ANs, number $M$ of TNs, communication range $R$, population size $L$, scaling factor $\alpha$, crossover probability $p_C$, maximum number $G$ of generations, lower bounds $a_1$ and $a_2$, and upper bounds $b_1$ and $b_2$ }
    \Output{ Estimated positions of the $M$ TNs}
	Initialize the populations according to (\ref{eq_Ini_pop}) and  (\ref{eq_Ini_pop_oppo})\;
%
%
 	\While{$g<=G$}{
 Select the best individual from each population and set the sharing matrix
 $\hat{\bm{X}}^{g}= \left[ \hat{\bm{x}}_{N+1}^g, \hat{\bm{x}}_{N+2}^g, \cdots, \hat{\bm{x}}_{N+M}^g \right]$ 
 
 as the initial estimation of the $M$ TNs\;
    	\For{$m=1; m<=M; m=m+1$}{
    		
    		\For{$l=1; l<=L; l=l+1$}{
    		Create mutant individuals with (\ref{eq_Mutation})\;
    		Generate trial individuals via (\ref{eq_Crossover})\;
    		Obtain anchored individuals via (\ref{eq_Anchoring_general})\;
    		Redirect the anchored individuals according to  (\ref{eq_redi})\;
    		Select the best individual according to (\ref{eq_Selection})\;
  			}
  			Update the $m$-th vector in $\hat{\bm{X}}^{g}$\;
  		}
  		Increase $g=g+1$\;
  	}
  	Return the estimated positions of the $M$ TNs\;
    \caption{Pseudo code of the proposed algorithm to estimate the positions of $M$ TNs.}
\end{algorithm}


\section{Experimentation and Performance Analysis}

\label{sec_Simulation}
In this section, results from
numerical simulations
are presented to show the performance of the
proposed algorithms
for solving the cooperative localization problem.
We
assume
an area of $100$ m $\times$ $100$ m with nine fixed ANs located at $(0,0)$, $(0,50)$, $(0,100)$, $(50,100)$, $(50,50)$, $(50,0)$, $(100,100)$, $(100,50)$, and $(0,100)$.
The RSS measurements are generated using the propagation model (\ref{eq_PL}) with a path loss $L_0=40$ dB, path loss exponent $\gamma=3 $,
and the standard deviations of the log-shadowing noise fixed to be the same for both connections AN-TN and AN-TN, i.e., $\sigma _{v_{nm}}=\sigma$.
For the proposed algorithms,
we assume a population size $L=10$, scaling factor $\alpha = 0.9$, crossover probability $p_{C}=0.9 $, lower bounds $a_1=a_2=0$, upper bounds $b_1=b_2=100$, and maximum number of generation $G=50$.
It
should be mentioned that
these values
were obtained after extensive preliminary experiments and
are in line with those used in \cite{caceres2020fastDEOR,caceres2020DEOR}.

For comparing the performance of the proposed algorithms,
the performances of the
SDP \cite{jour_Tomic}, SDP1 \cite{jour_Z_Wang_2019}, and
SOCP \cite{jour_Chang_2018}
approaches are taken into consideration here.
The algorithms of these approaches
are all implemented
in Matlab.
For the algorithms based on SDP and SOCP,
we use the well-known package CVX \cite{book_Boyd}, where the solver is 
SeDuMi
\cite{web_Polik_sedumi}.
In addition to these algorithms,
we have included the solution of the ML (ML-true) obtained through the function slqnonlin of Matlab, for which
the real positions of the TNs was assumed as the initial guessing positions to provide the actual lower bound on the localization error.
The
function slqnonlin employs the Levenverg Marquart optimization to solve the problem at hand.

The normalized root mean square error (NRMSE) defined as
%
%
\begin{IEEEeqnarray}{rCl} \label {eq:NRMSE}
	{\rm{NRMSE}} = \sqrt {\frac{1}{M{M_C}}\sum\limits_{i = 1}^{{M_C}} {\sum\limits_{m = 1}^M {{{\left\| {{\bm{x}_{mi}} - {{\hat{\bm{x}}}_{mi}}} \right\|}}} } }
\end{IEEEeqnarray}
%
is used as the performance metric,
where ${M_C}$ is the number of Monte Carlo runs,
and $\bm{x}_{mi}$ and ${\hat{\bm{x}}}_{mi}$ denote
the real and estimated positions, respectively, of the $m$-th TN
at the $i$-th run.


\subsection{Influence of the Number of Target Nodes}
\begin{figure}[!t]
	\centering
	\includegraphics[width=\linewidth]
	{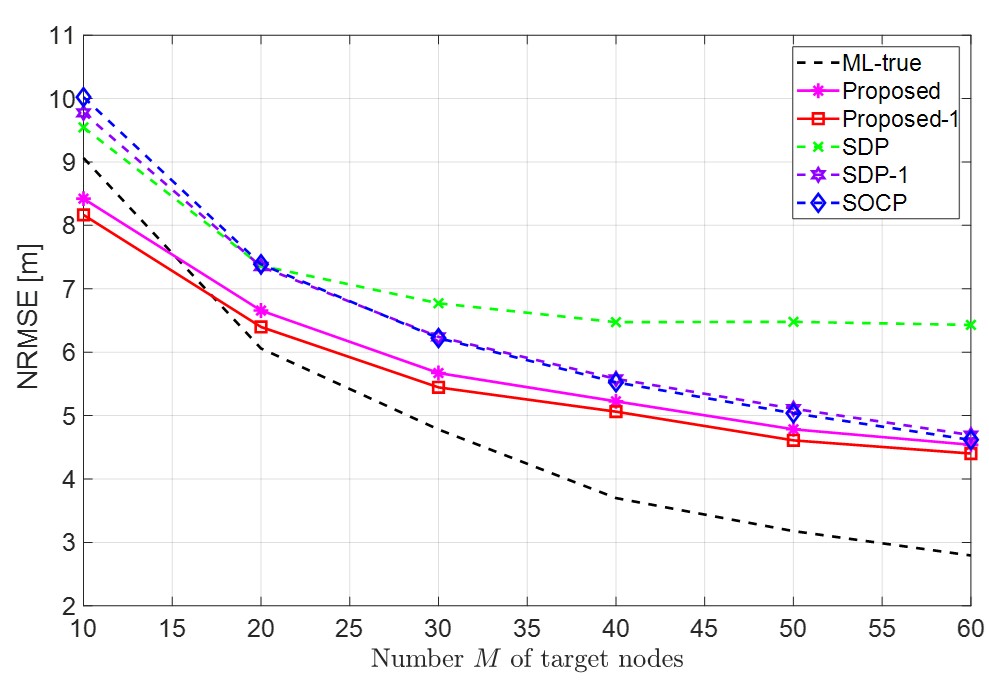} 
	\caption{NRMSE versus number of target nodes.}
	\label{fig_NRMSE_vs_Numb_TNs}
\end{figure}

The performance of the proposed algorithms is
first
evaluated
when the
number $M$ of TNs varies in the area of interest.
Fig. \ref{fig_NRMSE_vs_Numb_TNs} illustrates the NRMSE values as a function of the number of TNs when the connectivity range and standard deviation of the log-shadowing noise are set to $R=40$ m and $\sigma=4$ dB, respectively.
In general,
%
 it is observed that the localization accuracy of all algorithms improves as the number of TNs increases, as expected, due to the increased information from the TNs.
%
%
%
The proposed algorithms provide the best and most consistent localization accuracy, following the ML-true algorithm. Specifically, the proposed algorithms demonstrate a larger advantage over the SDP, SDP-1, and SOCP when the number of TNs is smaller. Additionally, the proposed algorithm using the population midpoint scheme achieves better localization accuracy than the proposed algorithm with the best individual selection among the two proposed algorithms.

\subsection{Influence of the Noise Variance}
\begin{figure}[!t]
\centering
  \includegraphics[width=\linewidth]
  {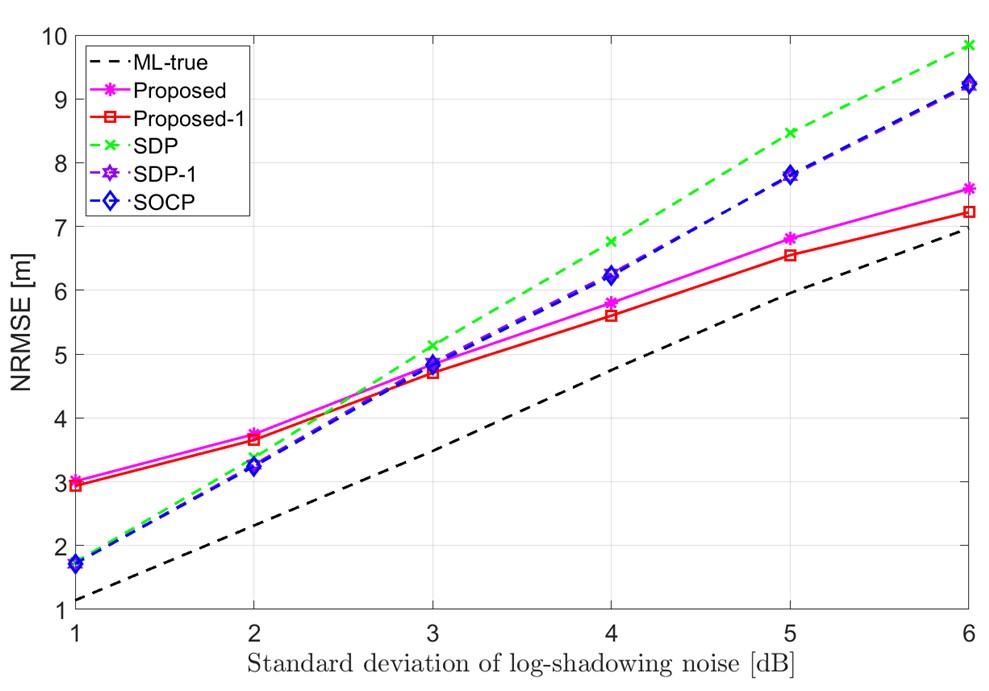}
  \caption{NRMSE versus standard deviation of the log-shadowing noise.}
\label{fig_NRMSE_vs_Noise}
\end{figure}

%
%
The performance of the algorithms is now evaluated
at
different levels of log-shadowing noise to investigate their effectiveness. Specifically, we considered a scenario with $M = 30$ TNs and a connectivity range of $R = 40$ m for all sensor nodes.
Fig \ref{fig_NRMSE_vs_Noise} shows the localization accuracy of the algorithms considered when the standard deviation of the log-shadowing noise varies from 1 to 6 dB.
%
As expected,
it is observed that
the performance of the algorithms
deteriorates
as the standard deviation increases.
However,
such trend is less prominent for the proposed algorithms.
%
This figure also shows that the proposed algorithms provide the least localization accuracy
for low values of the standard deviation of the log-shadowing noise,
the proposed algorithms outperform considerably the other algorithms at high values of the standard deviation of the noise.
For instance, at $\sigma=6$ dB, the proposed algorithm provides a localization improvement of around $23\%$, $18\%$, and $18\%$ with respect to the SDP, SDP1, and SOCP, respectively.
Similarly,
the enhanced proposed algorithm (Proposed-1) improves the localization accuracy by
roughly $26\%$, $22\%$, and $22\%$ with respect to the SDP, SDP1, and SOCP, respectively.

\subsection{Influence of the Connectivity Range}


Cooperative localization is a crucial aspect of
WSNs
and requires careful consideration of several parameters, including the connectivity range. However, increasing the connectivity range can lead to higher energy consumption, which can significantly reduce the life span of sensor nodes.
Consequently,
investigating
the localization accuracy as a function of the connectivity range is of great important.

\begin{figure}[!t]
	\centering
	\includegraphics[width=\linewidth]
	{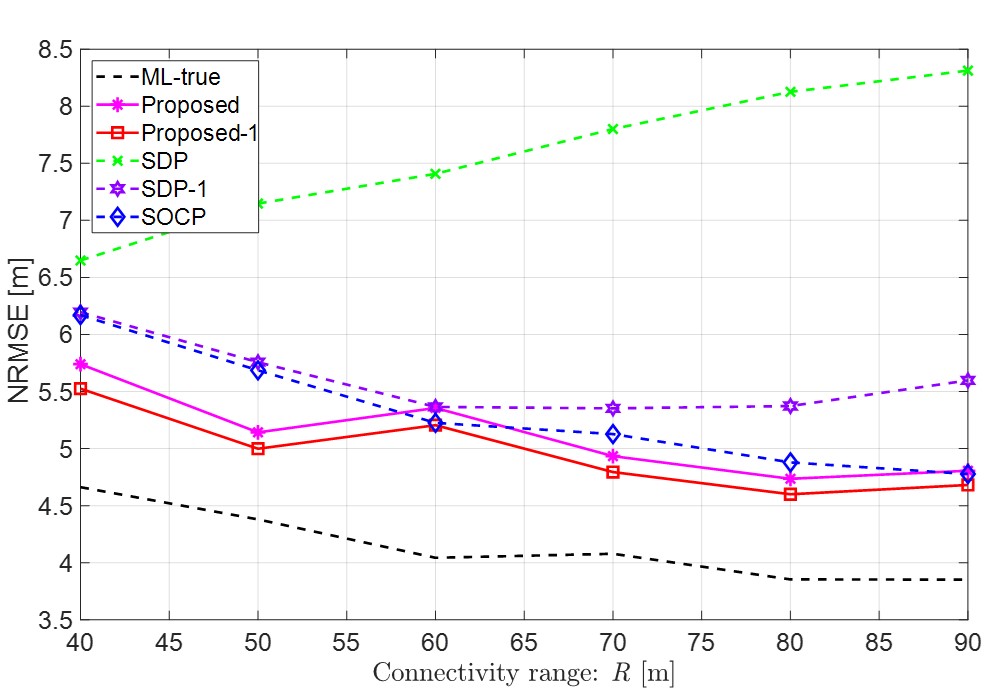} 
	\caption{NRMSE versus connectivity range.}
	\label{fig:NRMSE_vs_R}
\end{figure}

To investigate the impact of the connectivity range, we consider 30 TNs and a standard deviation $\sigma=4$ dB of the log-shadowing noise. Fig. \ref{fig:NRMSE_vs_R} shows the NRMSE
when
the connectivity range $R$
varies
from 40 to 90 m.
It is observed that, as $R$ increases, the NRMSE decreases for all the algorithms.
This is because a larger $R$ allows more connections
among
sensor nodes, increasing the number of RSS measurements.
Interestingly, the proposed algorithms provide the best performance when the connectivity range is smaller and have comparable localization accuracy at higher values of the connectivity range. This suggests that the proposed algorithms using connectivity information have a higher impact when the connectivity range values are smaller.


%

\subsection{Computational Complexity}
\label{sec_Complexity}
%
%
In cooperative localization, the computational complexity of an algorithm is also a crucial factor that should be taken into account alongside the localization accuracy. In fact, there exists often a trade-off between these two, and a balance needs to be targeted between achieving a higher localization accuracy and lower computational complexity.
An algorithm with high computational complexity may not be suitable for real-time processing and can limit its practical applicability. Conversely, reducing the computational complexity of an algorithm can enhance its applicability, particularly in real-time scenarios.
Thus, in this section, we examine and compare the computational complexity of the algorithms under consideration.

The \textit{Big-O} notation is used to describe the theoretical complexity of the algorithms.
In the calculation of the complexity of the algorithms,
minor terms are typically neglected, and only the dominant terms are considered.
In our particular case, the dominant terms are the maximum number $G$ of iterations, the number $L$ of individuals per population, the number $M$ of TNs, and the number $N$ of ANs.
Table \ref{tbl_Comp_Complexity}  presents the theoretical computational complexity for the algorithms discussed in the previous section.
%
%
%
%
It is noteworthy that the theoretical complexity of the proposed algorithms have an order of $N$ and $M^2$ with regards to the number $N$ of ANs and number $M$ of TNs, respectively.
%
In contrast,
the other algorithms used for comparison are of the order of
$N^2$ and $M^{6.5}$.
Thus, it is expected that the proposed algorithms will have a lower running computational time, especially when the numbers of TNs and/or ANs are high. This statement will be demonstrated shortly.

\begin{table}[!t]
	\centering
	\caption{Comparison of Theoretical Computational Complexity of Several Algorithms} 
	\begin{tabular}
		{l c}
		\toprule
		Algorithm & Complexity \\
		\toprule
		Proposed  & $O\left( 2GL \left( M^2+M \left(N-1\right) \right)\right)$ \\
		\midrule
		SDP \cite{jour_Tomic} & $O \left( \sqrt{2} M^{0.5} \left(4M^4 \left( N+\frac{M}{2}\right)^2 \right) \right)$ \\
		\midrule
		SDP1 \cite{jour_Z_Wang_2019} &  $O \left( M^{0.5} \left(4M^4 \left( N+\frac{M}{2}\right)^2 \right) \right)$ \\
		\midrule
		SOCP \cite{jour_Chang_2018} &  $O \left( M^{0.5} \left(4M^4 \left( N+\frac{M}{2}\right)^2 \right) \right)$ \\
		\bottomrule
	\end{tabular}%
	\label{tbl_Comp_Complexity}%
\end{table}%

\begin{figure}[!t]
	\centering
	\includegraphics[width=\linewidth]
	{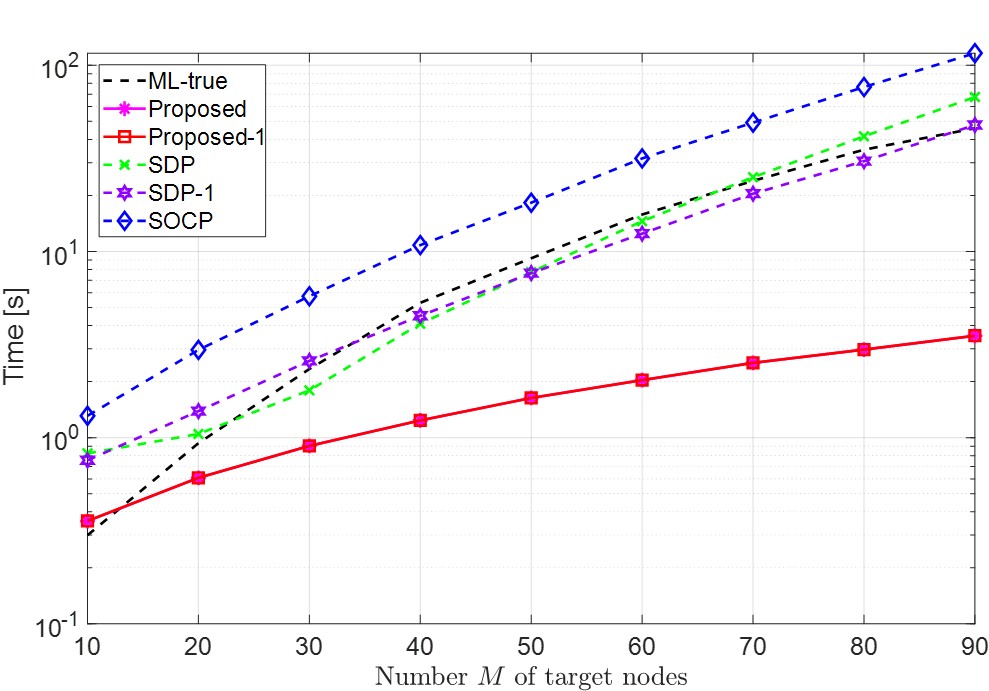}
	\caption{Average running time versus number of
		target nodes (CPU: Intel (R) Core (TM) i5-6600 3.30 GHz. RAM: 16.0 GB.).}
	\label{fig_SimuTimevsANs}
\end{figure}

Assuming $R=40$ m, $G=50$, $L=10$, $N=9$, and varying the number $M$ of TNs from $10$ to $90$, measurements of the real computational time are secured.
Fig. \ref{fig_SimuTimevsANs} shows the average of the computational time of all the algorithms in a log scale.
%
This figure
%
shows the proposed algorithms are the least demanding among all the algorithms compared.
%
This result
is in accordance with the theoretical complexity presented in Table \ref{tbl_Comp_Complexity}.
Additionally,
the figure also shows that both
proposed
algorithms have practically the same computational time.


\subsection{Discussion}
The comparison of the proposed algorithms have been performed with several variations of the cooperative localization scenario such as number of TNs, standard deviation of the log-shadowing noise, and connectivity range.
In such variation of the cooperative localization problem, the proposed algorithms provide significant improvement of the localization accuracy under high level of noise, limited connectivity range, and small number of TNs.
These results suggest the proposed algorithms are well suited for sparse deployment of WSNs.
In WSNs with a large number of sensor nodes, the proposed algorithms are expected to provide equal or only marginal gain in the localization accuracy with respect to the other algorithms, especially with respect to
those
based on the SOCP.
%
%
%
%
%

With regard to
the computational complexity,
we demonstrated,
theoretically and experimentally,
that the proposed algorithms require the least computational complexity.
%
This result implies
a better trade-off between computational complexity and accuracy of the proposed algorithms.
Additionally,
it is worth mentioning
that the proposed algorithms provide similar computational complexity
in
the localization accuracy gain obtained by incorporating the population midpoint scheme comes at free cost.
Thus, a proper incorporation of the population midpoint scheme in any evolutionary algorithm is highly recommendable.

%
%
%
\section{Conclusion}
\label{sec_Conclusion}

In this paper,
we developed an algorithm and an enhanced version of it based on differential evolution with opposition based learning, adapting redirection, anchoring, and population midpoint scheme.
The proposed algorithms are for addressing the high non-linearity, nonconvexity, and multi-modality of the cooperative localization problem in wireless sensor networks with limited communication range.
We also reformulated the problem under the multi-objective framework, where the cost function of the cooperative localization is split into several small cost functions that accounts only for a target node.
The proposed algorithms, which employ as many populations as the number of target nodes, have been compared with state-of-the-art algorithms based on semidefinite programming and second order cone programming.
%
%
Results from various simulations conducted under different scenarios indicate that the proposed algorithms offer a notable enhancement in localization accuracy compared to alternative methods. This improvement is particularly evident in scenarios characterized by a limited number of anchor nodes, high levels of noise, and a restricted connectivity range. Importantly, the proposed algorithms demonstrate comparable localization accuracy to other methods in alternative scenarios.
Additionally,
the computational complexity of the proposed algorithms is the least demanding among the algorithms compared.
%
%
These results imply the robustness and computational efficiency of the proposed algorithms
for sparse wireless sensor networks applications with limited communication range and computational power.

\section*{Acknowledgement}
This research was supported in part by the Basic Science Research Program through the National Research Foundation of Korea (NRF), funded by the Ministry of Education under
the grant numbers NRF-2020R1A4A1018774 and NRF-2021M3A9E4080780, 
and
%
by the
Program of International Visiting Professor of the University of Science
and Technology of China under Grant 2022BVT04, and by 
the research fund from Chosun University, 2024.



\bibliographystyle{IEEEtran}
\bibliography{RefBibioReview.bib}

\end{document}